\documentclass[aps,prd,11pt,preprint,nofootinbib,showkeys,longbibliograohy]{revtex4-2}
\usepackage{amsmath,amssymb,amsthm,mathrsfs,bbm}
\usepackage{latexsym,amscd,amsbsy,amsfonts,dsfont}
\usepackage{graphicx}
\usepackage[utf8]{inputenc}
\usepackage{subfigure}  
\usepackage{float}
\usepackage{slashed}
\usepackage{cancel}
\usepackage{multirow}
\usepackage{soul}
\usepackage{physics}
\usepackage{comment}
\usepackage{enumitem}
\usepackage{bm}
\usepackage{ulem}
\usepackage[usenames,dvipsnames]{xcolor}
\usepackage[
      colorlinks=true,
      linktocpage=true,
      breaklinks=true,
      linkcolor=Aquamarine,
      urlcolor=Purple,
      filecolor=black,
      citecolor=Purple,
      menucolor=black,
      pdfstartview=FitV,
      bookmarksopen=true
      ]{hyperref}

\newcommand*\xbar[1]{%
  \hbox{%
    \vbox{%
      \hrule height 0.5pt 
      \kern0.4ex
      \hbox{%
        \kern-0.1em
        \ensuremath{#1}%
        \kern-0.1em
      }%
    }%
  }%
} 

\usepackage{enumitem}
\setlist{noitemsep}

\makeindex

\begin{document}

\title{Cone hierarchy and the screening of matter by gravity}

\author{Julio Arrechea}
\email{julio.arrechea@sissa.it}
\affiliation{SISSA, via Bonomea 265, 34136 Trieste, Italy}
\affiliation{
INFN (Sez. Trieste), via Valerio 2, 34127 Trieste, Italy.}
\affiliation{IFPU, via Beirut 2, 34014 Trieste, Italy.}
\author{Carlos Barcel\'o}
\email{carlos@iaa.es}
\affiliation{Instituto de Astrof\'{\i}sica de Andaluc\'{\i}a (IAA-CSIC), Glorieta de la Astronom\'{\i}a, 18008 Granada, Spain}
\author{Gil Jannes}
\email{gil.jannes@ucm.es}
\affiliation{Department of Financial and Actuarial Economics \& Statistics, Universidad Complutense de Madrid, Campus Somosaguas s/n, 28223 Pozuelo de Alarc\'on (Madrid), Spain}

\begin{abstract}
In a previous paper by some of the authors (Gen. Rel. Grav. 56, 116, 2024), we  
introduced a novel paradigm with which to understand gravitational phenomena. We called it the Harmonic Background Paradigm (HBP).
In this paradigm, gravity amounts to an effective causality deformation with respect to a more fundamental causality, which always encompasses the former through a causal cone hierarchy. In that paper, the fundamental idea was described in detail but fully elaborated only when restricted to the linear gravitational approximation.
In this work, we discuss and conjecture how this idea could be extended to the full non-linear regime. We identify a connection between the 
cone hierarchy and a property of gravity that can be described as a screening mechanism of negative-energy gravitational clouds surrounding (but never overcoming) positive-energy seeds. 
We illustrate our ideas by applying them to spherically symmetric matter distributions. The paper concludes with a discussion of some key implications and directions for future research, including some remarks beyond General Relativity.
\end{abstract}



\maketitle
 
\tableofcontents

\section{Introduction}
\label{Sec:Introduction}

It is well known that gravity can be understood equivalently from two different perspectives or paradigms. 
One is the Geometric Paradigm based on a manifold with metric $g_{\mu\nu}$, and the other one is the Field-Theoretic Paradigm where gravity is described as a non-linear gauge theory of a tensor $h_{\mu\nu}$ on top of a Minkowski background metric $\eta_{\mu\nu}$.
The formulation of the latter paradigm is bi-metric, but the actual relevance of the background metric $\eta_{\mu\nu}$ is almost inconsequential. In particular, the causality of the $\eta$-metric does not play any role. For instance, depending on the selected gauge, the causal cones of the full metric $g_{\mu\nu}=\eta_{\mu\nu}+h_{\mu\nu}$ can be outside or inside those of the $\eta$-metric~\cite{GaoWald2000}.
Another way to describe this problem is that there is no canonical way to associate points in the effective metric manifold $\{{\cal M},g\}$ with points in the background metric manifold $\{\cal M,\eta\}$ (see e.g. \cite{Lenzi:2021wpc}).

In a previous paper~\cite{BarceloJannes2024}, we have proposed a third paradigm, distinct from the previous two standard paradigms, to understand gravitational phenomena. We named it the Harmonic Background Paradigm (HBP). Its essence lies in considering the gravitational field as a tensor $h_{\mu\nu}$ subject to algebraic constraints plus boundary/initial conditions. Crucially, the dynamical theory satisfied by this object does not contain any gauge symmetry. In weak-gravity circumstances, this formulation is equivalent to the standard ones, leading to the same phenomenology. But even in this weak regime, we have argued that the HBP suggests compelling and novel interpretations of gravitational phenomena, such as a plausible explanation as to why gravity is attractive. Moreover, in the strong gravity regime, it hints towards specific new physical processes that might help to solve the main conceptual problem of general relativity: the development of singularities under gravitational collapse. 
Our HBP has ties with other attempts in the literature to construct special relativistic theories of gravitation~\cite{Rosen1940a,Rosen1940b,LogunovMestvirishvili1985,pitts2001slightly,PittsSchieve2004}.

In our previous work, the discussion of the HBP was restricted to the linear approximation. The central point of the construction is that in the HBP formulation, there must be a causal cone hierarchy between the causal cones provided by any $g=\eta+h$ metric and those of the background $\eta$ metric: the effective causal cones associated to $g$ are always inside the background causal cones. Formally, we write $[g]<[\eta]$. In our previous paper~\cite{BarceloJannes2024}, we were able to show the consistency of this proposal, but only at the linear level.

In this work we discuss which kind of additional requirements would be necessary for gravity to be described as a deformed causality always inside a more rigid background causality, also beyond the linear regime. We conjecture a specific construction and explain the ideas behind its potential correctness. In addition, we illustrate its most basic characteristics by applying it to spherically symmetric matter distributions.

The paper is structured as follows. Section~\ref{Sec:linear} provides a concise review of how the central idea emerges within the linear regime. Section \ref{Sec:HBP} then extends the Harmonic Background Paradigm to the non-linear regime. Our central conjecture in this extension is that the simplest condition for the construction to be consistent is to impose a condition on $h_{\mu\nu}$ which we will call the de Donder Harmonic Condition (dDHC). 
Subsequently, Section \ref{Sec:stars} demonstrates how spherically symmetric stars are represented within this paradigm. The paper concludes with a discussion of some key implications and directions for future research.

\section{Main idea at the linear level}
\label{Sec:linear}

The Fierz-Pauli action for linear gravity in Minkowski spacetime can be written:
\begin{align}
    & S = \int d^4 x \sqrt{ - \eta} \left( - \frac{1}{4} \nabla_{\mu} h^{\alpha \beta} \nabla^{\mu} h_{\alpha \beta} + \frac{1}{2} \nabla_{\mu} h^{\mu \alpha} \nabla_{\nu} h^{\nu}_{\alpha}
- \frac{1}{2} \nabla_{\mu} h^{\mu \nu} \nabla_{\nu} h + \frac{1}{4} \nabla_{\mu} h \nabla^{\mu} h \right).
\label{h_action}
\end{align}
Here, $\nabla$ represents the covariant derivative with respect to the flat metric $\eta_{\mu\nu}$ of Minkoswki spacetime. We use $\nabla$ instead of $\partial$ (as in the VBL paper) to stress the covariance of the construction under changes of coordinates.
From the Fierz-Pauli action, one can derive the field equations
\begin{align}
\frac{1}{2} \Box h_{\mu \nu} - \eta_{\beta ( \mu} \nabla_{\nu)} \nabla_{\alpha} h^{\alpha \beta} + \frac{1}{2} \left( \nabla_{\mu} \nabla_{\nu} h + \eta_{\mu \nu} \nabla^2 \cdot h \right)
- \frac{1}{2} \eta_{\mu \nu} \Box h  =  - 8\pi G \, T^{\rm M}_{\mu \nu}\,,
\label{h_equations}
\end{align}
where we have introduced the notation $\nabla^2 \cdot h = \nabla_{\mu} \nabla_{\nu} h^{\mu \nu}$ (which we will keep using throughout this section). We have added a stress-energy-tensor (SET) matter source on the right-hand side. It is easy to see that when taking a covariant derivative $\nabla_\mu$ and contracting, all the $h_{..}$ terms disappear. Thus, for consistency, the SET must be conserved:
\begin{equation}
    \nabla^\mu T^{\rm M}_{\mu \nu}=0.
\label{SETconservation}
\end{equation}

We can alternatively use a different gravitational variable, the trace-reversed perturbation
\begin{equation}
    \bar{h}^{\mu \nu} = h^{\mu \nu} - \frac{1}{2} \eta^{\mu \nu} h. 
\label{weyl_transf}
\end{equation}
In terms of this variable $\bar{h}^{\mu \nu}$, the Fierz-Pauli action reads:
\begin{align}
    S = \int d^4 x \sqrt{ - \eta} \left[ - \frac{1}{4} \nabla_{\mu} \bar{h}^{\alpha \beta} \nabla^{\mu} \bar{h}_{\alpha \beta} + \frac{1}{2} \nabla_{\mu} \bar{h}^{\mu \alpha} \nabla_{\nu} \bar{h}^{\nu}_{\alpha}+ \frac{1}{8} \nabla_{\mu} \bar{h} \nabla^{\mu} \bar{h}  \right],
\label{barh_action}
\end{align}
and the field equations become
\begin{align}
\frac{1}{2} \Box \bar{h}_{\mu \nu} - \eta_{\beta ( \mu} \nabla_{\nu)} \nabla_{\alpha} \bar{h}^{\alpha \beta} 
 - \frac{1}{4} \eta_{\mu \nu} \Box \bar{h}  =  - 8\pi G \,\bar{T}^{\rm M}_{\mu \nu},
\label{barh_equations}
\end{align}
where
\begin{equation}
    \bar{T}^{\rm M}_{\mu \nu} = T^{\rm M}_{\mu \nu} - \frac{1}{2} \eta_{\mu \nu} T^{\rm M},~~~~~T^{\rm M}=\eta^{\mu \nu}T^{\rm M}_{\mu \nu}.
\label{weyl_transf}
\end{equation}
The field equations can yet equivalently be written as
\begin{align}
\frac{1}{2} \Box \bar{h}_{\mu \nu} - \eta_{\beta ( \mu} \nabla_{\nu)} \nabla_{\alpha} \bar{h}^{\alpha \beta} + \frac{1}{2} \eta_{\mu \nu} \nabla^2 \cdot \bar{h} =  - 8\pi G \, T^{\rm M}_{\mu \nu}.
\label{h_equations}
\end{align}
Now comes the idea first proposed in~\cite{VisserBassettLiberati2000}, further referred to as the VBL paper, and further elaborated by the present authors in~\cite{BarceloJannes2024}. Consider the case in which: 
\begin{enumerate}
\item 
The actual gravitational field is the tensor $h_{\mu\nu}$ but subject to the de Donder Harmonic Condition (dDHC) $\nabla^\mu \bar{h}_{\mu\nu}=0$.
\item 
The initial configuration for $h_{\mu\nu}$ is such it does not contain incoming waves at past infinity (neither actual physical waves nor gauge waves, as detailed in ~\cite{BarceloJannes2024}). 
\item 
The matter SET satisfies a Background Null Energy Condition (B-NEC), namely that $T_{\mu\nu} l^\mu l^\nu \geq 0$ for any future-pointing null vector $l^\mu$ with respect to $\eta$, i.e.\ any $l^\mu$ such that $\eta_{\mu\nu}l^\mu l^\nu=0$.
\end{enumerate}
Then, one first realizes that the equation of motion of the field is simply
\begin{align}
\Box \bar{h}_{\mu \nu} =  - 16 \pi G \, T^{\rm M}_{\mu \nu}~;~~~{\rm or}~~~~\square h_{\mu\nu} = -16\pi G \bar{T}^{\rm M}_{\mu\nu}.
\label{h_equations}
\end{align}
Under condition 2.\ above, one can write the evolved form of $h_{\mu\nu}$ as
\begin{eqnarray}
h_{\mu\nu}(\bm{x},t)= 4\int d \bm{x}' \frac{\bar{T}_{\mu\nu}(\bm{x}',t')}{\abs{\bm{x}-\bm{x}'}}.
\label{VBL-int}
\end{eqnarray}
$\bar{T}_{\mu\nu}$ in the above equation depends on $\bm{x}'$ and on the retarded time $t'=t-\abs{\bm{x}-\bm{x}'}$ (we use $c=1$ units).

Then, under the previous conditions, it is easy to see that the causal cones of the effective metric $g_{\mu\nu}=\eta_{\mu\nu}+h_{\mu\nu}$ will always lie inside the cones of the $\eta$ metric. Indeed, contracting \eqref{VBL-int} with $l^\mu l^\nu$ immediately gives
\begin{equation}
h_{\mu\nu} l^\mu l^\nu \geq 0.
\label{VBL-null}
\end{equation}
But since by definition $l^\mu$ is null with respect to $\eta$, the expression \eqref{VBL-null} is also the norm of $l^\mu$ with respect to the full metric $g$. Hence, $l^\mu$ cannot be timelike with respect to $g$, and in fact, unless $T_{\mu\nu}\equiv 0$, it will necessarily be spacelike with respect to $g$. Thus, the causal cones of $g$ always lie inside the causal cones of $\eta$, which we indicate by writing $[g]<[\eta]$. 

To summarize the VBL argument: the B-NEC automatically implies that the $g$-lightcones will be squeezed inside the $\eta$-lightcones. Note the crucial role played by the B-NEC, not the standard NEC, since the latter would involve null vectors with respect to $g$ rather than $\eta$. 

Another important point to note here is that, for $h_{\mu\nu} l^\mu l^\nu$ to be positive at one point, it is not strictly necessary that $T_{\mu\nu} l^\mu l^\nu$ is always and everywhere positive. It is only necessary that any integration along the past causal cone of that point gives a positive result. And this should be true for any point of the manifold. Essentially, the idea is that the gravitational potential field $h_{\mu\nu}$ at any particular spacetime point is picking up retarded contributions from all the matter sources in the universe that are causally connected to it. If the addition of all these contributions is positive, the causal cones of the $g$-metric at that point will lie inside those of the $\eta$-metric.

The previous argument is strictly valid in the linear case. As explained in detail in \cite{BarceloJannes2024}, it remains valid in spite of the gauge-related Gao-Wald criticism \cite{GaoWald2000} when interpreted within the HBP and, in fact, provides an elegant explanation for the attractive character of gravity in all standard weak-field situations. However, in the previous papers \cite{VisserBassettLiberati2000} and \cite{BarceloJannes2024}, there is no proof nor argument that the same hierarchy $[g]<[\eta]$ should be valid at the non-linear level. We will now provide an argument in defense of this idea.

\section{The Harmonic Background Paradigm}
\label{Sec:HBP}

\subsection{Non-linear equations}
\label{Subsec:Non-linear}

The Einstein-Hilbert Lagrangian can be written as 
\begin{eqnarray}
\mathcal{L}_{\text{EH}} = \frac{1}{8\pi G}\sqrt{-g} \, g^{\mu\nu} g^{\lambda\sigma}\left[
\partial_\lambda \Gamma_{\sigma,\mu\nu} 
- \partial_\nu \Gamma_{\sigma,\mu\lambda} 
+ g^{\rho\epsilon}\left(\Gamma_{\sigma,\lambda\rho} \Gamma_{\epsilon,\mu\nu} 
- \Gamma_{\sigma,\nu\rho} \Gamma_{\epsilon,\mu\lambda} 
\right)
\right].
\end{eqnarray}
This can be shown to be equivalent to
\begin{eqnarray}
\mathcal{L}_{\text{EH}} = \frac{1}{8\pi G}\sqrt{-\eta} \, {\sqrt{-g} \over \sqrt{-\eta}} g^{\mu\nu} g^{\lambda\sigma}\left[
\nabla_\lambda C_{\sigma,\mu\nu} 
- \nabla_\nu C_{\sigma,\mu\lambda} 
+ g^{\rho\epsilon}\left(C_{\sigma,\lambda\rho} C_{\epsilon,\mu\nu} 
- C_{\sigma,\nu\rho} C_{\epsilon,\mu\lambda} 
\right)
\right],
\end{eqnarray}
where $C_{\sigma,\mu\lambda}$ has the same structure as $\Gamma_{\sigma,\mu\lambda}$ but with $\partial_\mu$ substituted by covariant derivatives $\nabla_\mu$ with respect to the background $\eta$-metric. 
It is interesting to recall that $C$ is now a tensorial object, as opposed to the non-tensorial character of $\Gamma$.
Integrating by parts and discarding surface terms, we find another equivalent expression for the gravitational Lagrangian $\mathcal{L}_{\rm G}$:
\begin{eqnarray}
\mathcal{L}_{\rm G} = -
\frac{1}{8\pi G}\sqrt{-\eta} \, {\sqrt{-g} \over \sqrt{-\eta}} g^{\mu\nu} g^{\lambda\sigma} g^{\rho\epsilon} \left( 
C_{\sigma,\mu\nu} C_{\epsilon,\lambda\rho} 
- C_{\sigma,\mu\rho} C_{\epsilon,\nu\lambda}
\right).
\end{eqnarray}
This is Rosen's Lagrangian~\cite{rosen1973bi,rosen1975bi}.

With some further manipulations, the Lagrangian could also be written as a (infinite) series of terms of the form $(h_{..})^m\nabla_{.} h_{..} \nabla_{.} h_{..}$ (contracted with a number of $\eta^{..}$), or equivalently of the form $(\bar{h}_{..})^m\nabla_{.} \bar{h}_{..} \nabla_{.} \bar{h}_{..}$. The key point is that this allows us to write the Einstein equations with a matter source as
\begin{align}
\frac{1}{2} \Box \bar{h}_{\mu \nu} =  \eta_{\beta ( \mu} \nabla_{\nu)} \nabla_{\alpha} \bar{h}^{\alpha \beta} - \frac{1}{2} \eta_{\mu \nu} \nabla^2 \cdot \bar{h} - \tilde{T}_{\mu\nu}^{\rm GNL}(h) - 8\pi G \, T^{\rm MI}_{\mu\nu}(h,\Phi) -8\pi G \, T^{\rm MF}_{\mu \nu}(\Phi).
\label{NLh_equations_1}
\end{align}
Since at the linear level, the gravitational field does not contribute to the total energy, we have denoted as a gravitational stress-energy contribution $\tilde{T}_{\mu\nu}^{\rm GNL}$ the gravitational non-linear (GNL) contributions. 
This SET contains terms of quadratic and higher order in $h_{\mu\nu}$. 
At this stage we do not multiply and divide this GNL term by $8\pi G$ to show explicitly that its coupling is different from that of the matter SET $T^{\rm M}_{\mu\nu}$. 
The $\Phi$ fields in this matter SET $T^{\rm M}_{\mu\nu}$ generically denote any type of matter content present, and we have separated $T^{\rm M}_{\mu\nu}$ into two parts:
\begin{align}
T^{\rm M}_{\mu\nu}(h,\Phi) = T^{\rm MI}_{\mu\nu}(h,\Phi) + T^{\rm MF}_{\mu \nu}(\Phi).
\label{DDHC}
\end{align}
The first part, $T_{\mu\nu}^{\rm MI}$, represents an interaction term between matter and gravity, whereas the second part, $T_{\mu\nu}^{\rm MF}$, contains all contributions that do not depend on $h_{\mu\nu}$ and so can be considered a free-matter term. 
Taking a $\nabla$ derivative and contracting the indices, it is clear that
\begin{align}
\nabla^\mu (\tilde{T}_{\mu\nu}^{\rm GNL} + 8\pi G \,T^{\rm MI}_{\mu\nu} + 8\pi G \, T^{\rm MF}_{\mu \nu})=0.
\label{total-conservation}
\end{align}
Therefore, the total SET is conserved, not as a covariant conservation with respect to the metric $g$, but as a strict conservation with respect to Minkowski observers.

An important observation is the following. 
The $\tilde{T}_{\mu\nu}^{\rm GNL}$ term on the right-hand side of Eq.~\eqref{NLh_equations_1} contains terms not only of the form $h_{..}^m \nabla_. h_{..} \nabla_. h_{..}$ but also of the form $h_{..}^m \nabla_. \nabla_. h_{..}$ (with $m\geqslant 0$). 
This is not surprising since $h_{..}$ is coupled non-minimally to $h_{..}$ itself. In terms of the characteristic evolution of the partial differential equation (PDE) problem, these non-linearities are changing the effective causality of the system from the Minkowskian one that shows up in the linear theory (with just the flat D'Alembertian), to a distorted one in the full non-linear theory. A well-known manner to correctly understand the causal properties of the PDE system is to impose the gauge condition~\cite{Foures-Bruhat:1952grw,Friedrich:1985afv,Friedrich:1996hq}
\begin{eqnarray}\label{Fock-condition}
    g^{\mu\nu}C_{\alpha,\mu\nu}=0.
\end{eqnarray}
This is simply the standard Fock Harmonic condition~\cite{Fock1959}. One can then verify that the only second-order terms remaining in the full Einstein equations are 
\begin{align}
g^{\alpha\beta}\nabla_\alpha \nabla_\beta g_{\mu\nu}~.
\label{Bruhat-condition}
\end{align}
From here one can immediately appreciate that the characteristics of the PDE system follow the causal structure defined by the $g$-metric. This causal structure is independent of which gauge condition one uses. 
Notice, however, that this causal analysis by itself does not tell us anything regarding the hierarchy between the $g$-metric and the $\eta$-metric. 

Coming back to Eq.~\eqref{NLh_equations_1}, let us restrict the form that $h_{..}$ can adopt by imposing the algebraic restriction 
\begin{align}
\eta^{\mu\nu}\nabla_{\mu} \bar{h}_{\nu\sigma}=0.
\label{DDHC}
\end{align}
This condition can equivalently be written as 
\begin{align}
\eta^{\mu\nu} C_{\sigma,\mu\nu}=0.
\label{DDHC-connection}
\end{align}
We will call it the ``de Donder Harmonic Condition" (dDHC). Notice that it is a straightforward extension from the linear case. However, it is crucially different from the standard (non-linear) Fock Harmonic condition~\eqref{Fock-condition} just mentioned in the context of the characteristics of the PDE system.

Imposing the dDHC, the Einstein equations now acquire the form
\begin{align}
\frac{1}{2} \Box \bar{h}_{\mu \nu} =  - \tilde{T}_{\mu\nu}^{\rm GH}(h) - 8\pi G \, T^{\rm MIH}_{\mu\nu}(h,\Phi) -8\pi G \, T^{\rm MF}_{\mu \nu}(\Phi),
\label{NLh_equations_2}
\end{align}
where GH and MIH represent the part of the gravitational non-linear term $\tilde{T}_{\mu\nu}^{\rm GNL}$ and the interaction term $T_{\mu\nu}^{\rm MI}$, respectively, that remain after explicitly imposing the dDHC. These expressions differ formally from the previous ones only in the sense that we eliminate terms that include $\nabla^\mu \bar{h}_{\mu\nu}$ factors. The conservation of the total stress-energy tensor with respect to the background metric $\eta$ is maintained:
\begin{align}
\nabla^\mu (\tilde{T}_{\mu\nu}^{\rm GH} + 8\pi G \, T^{\rm IH}_{\mu\nu} + 8\pi G \, T^{\rm MF}_{\mu \nu})=0.
\label{total-conservation}
\end{align}

Therefore, at the end of the day, we can write the same formal expression as in the linear case, see Eq.~\eqref{SETconservation}, but now considering a SET composed of three parts instead of only a free matter part 
(one just has to define $8\pi G T_{\mu\nu}^{\rm GH} = \tilde{T}_{\mu\nu}^{\rm GH}$ to set the three parts of the SET on an equal formal footing).

The key question is now: Is it possible to formulate sufficient conditions that the matter SET should obey in order to guarantee that the $g$-lightcones lie strictly inside the $\eta$-lightcones, just like in the linear case?
Notice first of all that our initial philosophy is to impose restrictions only on the matter sector. 
The gravity SET will then be interpreted as a reactive contribution which automatically adapts itself to 
the evolution of matter. Keeping this remark in mind, we will argue in the next section that such conditions indeed exist, and moreover are perfectly reasonable.

\subsection{Gravity screening matter}
\label{Subsec:screening}

Since we have a background $\eta$ metric, we can use the constant time slices associated with a particular inertial reference frame to describe the evolution. When looking at some initial time $t_0$, the structure of $h_{..}$ at $t_0$ could be very complex. In particular, the initial $h_{..}$ could be such that $[g] \nleq [\eta]$;  also, $h_{..}$ could be non-vanishing at infinity, pointing to a non-flat asymptotic structure. With such initial conditions, it proves difficult to demonstrate that the evolution will be such that $[g] \leq [\eta]$ at some later time $t_1$.

We therefore consider that initially the situation is ``well-behaved", in the sense that both $[g] \leq [\eta]$, and $h_{\mu\nu} \to 0$ at spatial infinity so that the $g$-metric is asymptotically flat. However, this by itself does not guarantee a dDHC nor an energy condition for matter. For example, matter could violate different energy conditions at times before $t_0$ but the $g$-cones happen to be inside the $\eta$-cones at $t_0$. 

To have a clear understanding and control of the behaviour of the matter/gravity system, it is helpful and sufficient to assume that in the past and until $t_0$, the system has been in the linear regime described in the previous section. Let us recall that this implies the following set of conditions, up to $t_0$:
\begin{enumerate}
    \item First of all, remember that gravity is encoded in a tensorial object $h_{..}$ constrained to satisfy the dDHC; 
    \item Gravity is initially very weak; among other things this implies that the free matter SET, being itself small, is however much bigger than the gravity and interaction SETs;
    \item The tensor $h_{..}$ dies off at spatial infinity so that the effective metric $g_{..}$ is asymptotically flat;
    \item There are no gravitational waves incoming from infinity;
    \item Finally, we impose that the free matter SET initially satisfies the B-NEC.
\end{enumerate}
Therefore, on the one hand, we know by construction that $h_{\mu\nu}$ is very small until $t_0$. Moreover, we have a precise definition of what it means for $h_{..}$ to be small, namely that the $h_{\mu\nu}$ components are much smaller than the corresponding $\eta_{\mu\nu}$ components (recall that $h_{\mu\nu}$ is a unique object that cannot be gauge transformed). 

On the other hand, we are under the conditions of the linear analysis described in Section \ref{Sec:linear}. Therefore, we know that the relation $[g] \leq [\eta]$ is maintained, and this is guaranteed as long as the B-NEC condition is satisfied. 
The remaining (and potentially hard) problem is to see what happens when the system evolves and eventually enters into a non-linear regime. In fact, this is a natural evolution followed by gravitational systems. For example, a dilute molecular cloud can start collapsing under its own weight, making gravity more and more prominent until the linearity conditions are violated.

To analyze this non-linear case, let us maintain the dDHC. Then, we formally have the same expression~(\ref{VBL-int}) but now using the total SET 
\begin{align}
T^{\rm T}_{\mu\nu}= T^{\rm GH}_{\mu\nu} 
+ 
T^{\rm IH}_{\mu\nu} 
+ 
T^{\rm MF}_{\mu \nu}.
\label{total-conservation}
\end{align}
Now, if at some point in the future $[g] > [\eta]$ (even if only for some particular null directions), this would imply that the total SET has violated the integrated B-NEC along the past causal cones of the background metric (i.e., the $\eta$ metric). Since we have imposed that the free matter SET satisfies the B-NEC, this would necessarily mean that the gravity (and interaction) terms of the total SET have overtaken the free matter SET and violated the integrated B-NEC.

However, this hypothetical phenomenon appears to contradict one of the most important results in standard general relativity: the positive energy theorem~\cite{schoen1979proof,schoen1981proof,witten1981new,parker1982witten,faddeev1982energy}.  
The positive energy theorem tells us that, under the following two conditions: 
\begin{enumerate}[label=\roman*),noitemsep, topsep=0pt, parsep=0pt, partopsep=0pt]
    \item asymptotic decay of $h_{..}$ so that the metric manifold $\{{\cal M},g\}$ is asymptotically flat,
    \item satisfaction of the Dominant Energy Condition (DEC) by the matter SET,
\end{enumerate}
it is impossible to construct configurations with negative asymptotic energies. 
This can be interpreted as implying that the aggregation of positive-energy matter can produce clouds of negative gravitational energy but this negative gravitational energy can never overcome the positive contribution of its classical matter seed. To give a hand-waving description of the physics underneath the theorem, we have used the term ``clouds of negative gravitational energy". In standard GR, this clouds are, however, delocalized. Within the HBP this clouds are, instead, locally defined. 

\begin{figure}
\begin{center}
\includegraphics[scale=0.4]{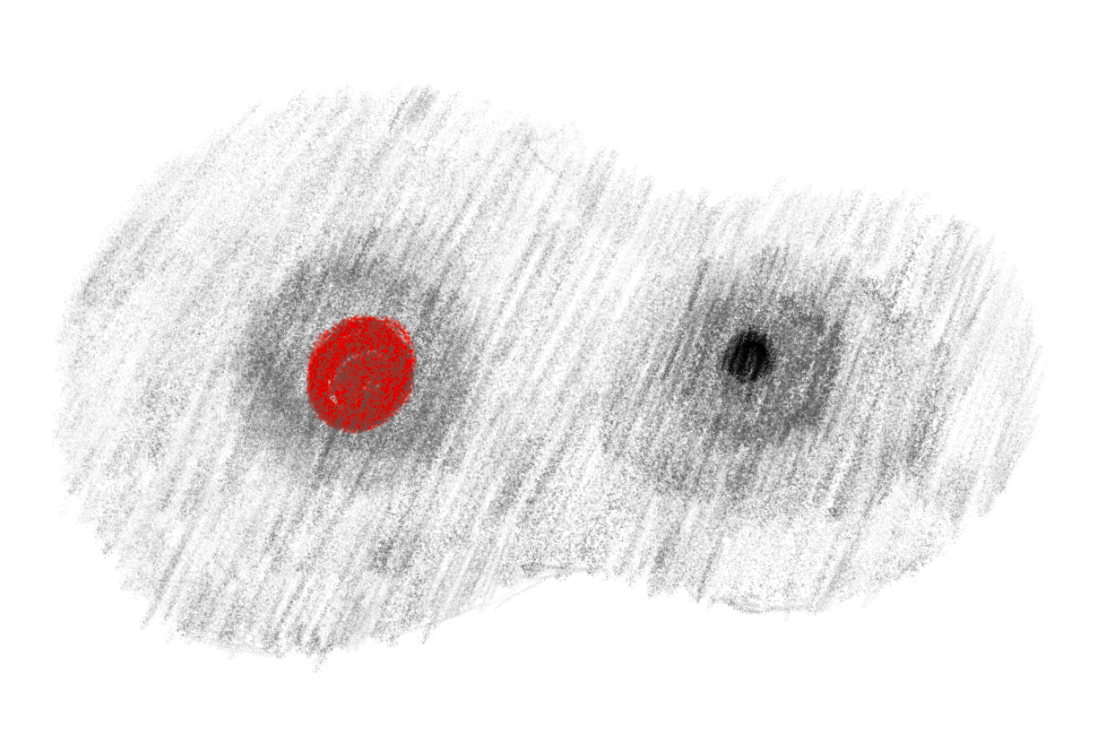}
\caption{Pictorial representation of two clouds of gravitational energy. In red we represent the presence of a positive-energy lump of matter. Such lumps of matter generate a cloud of negative gravitational energy, represented by the grey clouds. The standard case is on the left. The case in which the negative energy clouds completely overshadow the central positive matter contribution, and the overall net energy is thus negative, is depicted on the right. This would thus constitute a region of spacetime with repelling properties. This is precisely the type of configurations which we conjecture to be forbidden, through some non-trivial local version of the positive energy theorem.}
\label{Fig:GravitationalCloudsB}
\end{center}
\end{figure}

\begin{figure}
\begin{center}
\includegraphics[scale=0.4]{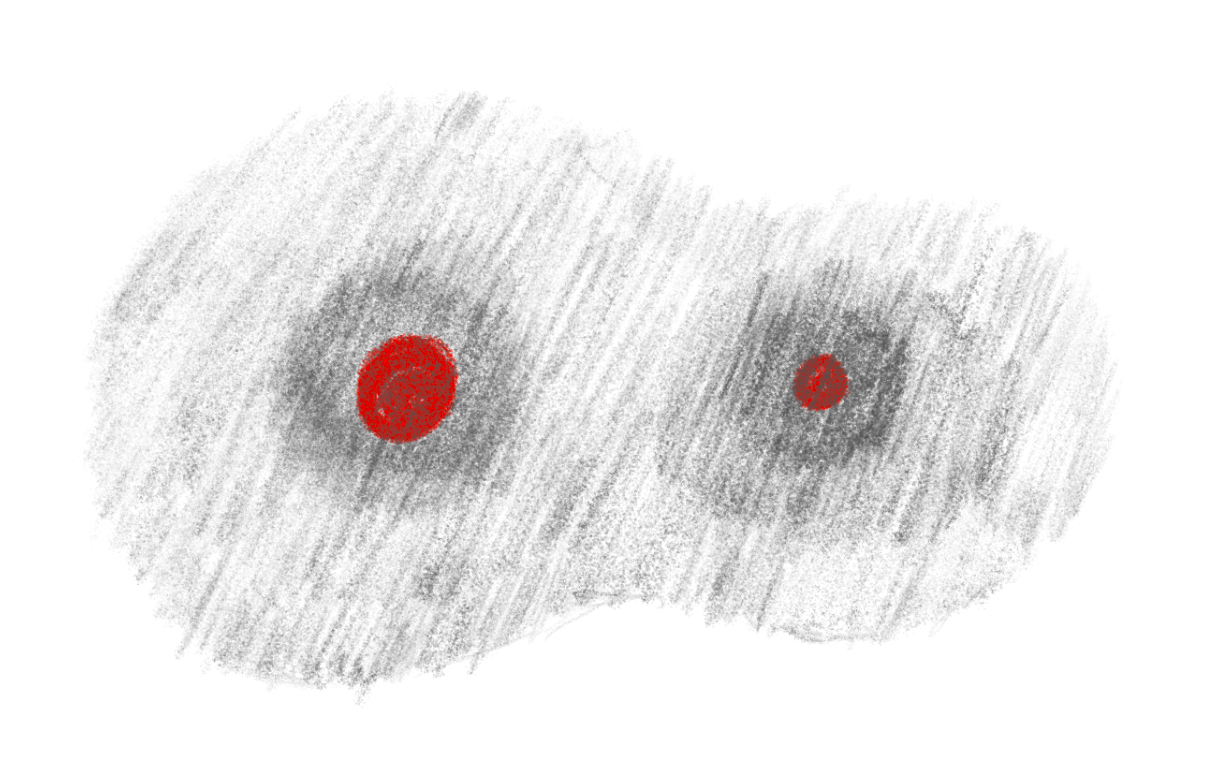}
\caption{Pictorial representation of how clouds of negative gravitational energy (in grey) accumulate surrounding matter seeds of positive energy (in red). Our conjecture is that the negative clouds never overshadow the central positive-energy matter seeds.}
\label{Fig:GravitationalCloudsA}
\end{center}
\end{figure}

A corollary relevant here is that no gravitational evolution seeded by DEC-satisfying matter can lead to isolated accumulations of negative energy (see Figure~\ref{Fig:GravitationalCloudsB}). If this were not true, then there should exist vacuum solutions of general relativity with asymptotic negative energy. But then Minkowski spacetime would not be the state of minimum energy and, in fact, the theory would be completely unstable. 
Thus, under the conditions of the positive energy theorem, general relativistic evolution can only generate negative gravitational energies surrounding (or ``dressing") the (positive-energy) matter concentrations (see Figure~\ref{Fig:GravitationalCloudsA}).

Unfortunately, the known theorems on energy positivity in standard GR (i.e., either within the geometrical or the field-theoretical paradigms) cannot be straightforwardly applied to our description. On the one hand, the Schoen-Yau-Witten theorem~\cite{schoen1979proof,schoen1981proof,witten1981new} is global so it says nothing about local gravitational energies, which within the standard paradigms make no sense. On the other hand, there exist other positive-energy theorems that use quasi-local notions of energy. Among several proposals (without being exhaustive, see~\cite{Komar:1963svp,Brown:1992br,Liu:2003bx}), the notion of quasi-local energy that has proved to have a better behaviour is that of Wang-Yau~\cite{WangYau2009}. They demonstrated that if the DEC is satisfied, their notion of quasi-local energy is always non-negative. 
The application of this result to our situation, however, is also not direct. Wang and Yau associate an energy content to a spatially bounded region of spacetime (a quasi-local notion). Here, on the other hand, we are implementing a strictly local definition of gravitational energy. Interestingly, they showed that in the ultralocal or small-ball limit, their energy definition matches the Bel-Robinson energy~\cite{Bel1959,Bonilla:1997ink}. The Bel-Robinson object is a tensorial quantity that contains up to fourth-order derivatives of the metric. At this stage we do not know whether our second-order derivatives expression for a local energy has some relation with the Bel-Robinson tensor, but it would be interesting to investigate this potential connection (for a related investigation, see~\cite{Deser:1999me}).

Another challenge with the application of the positive energy theorems to our problem here is the use of a different energy condition.
The standard DEC asserts that for any future-pointing causal vector $Y^\mu$ of the $g$-metric, the contraction $(T^{\rm M})^\mu_{~~\nu}Y^\nu$ is another future-pointing causal vector of the $g$-metric. Physically, it excludes the possibility of having fluxes of energy travelling outside the causal $g$-cone. Assuming for the moment the consistency assumption that $[g]<[\eta]$ (this is what we want to show), we can state the following. If the matter SET satisfies the DEC (defined with respect to $g$), then it will necessarily also satisfy a ``Background DEC" or B-DEC, defined with respect to $\eta$. Note that the reverse is not true: one could set up flows of energy that are causal from the point of view of the $\eta$ metric but violate the $g$-causality. Furthermore, if the matter SET satisfies the B-DEC, then it will automatically also satisfy the B-NEC, essential in the VBL argument~\cite{VisserBassettLiberati2000}.

Now, it is clear that our total SET will be negative in local vacuum regions (i.e., regions outside matter sources). Therefore, the total SET will violate the B-NEC locally in such regions, since its gravitational SET will be negative while $T^{\rm M}_{\mu\nu}=0$. Thus, our conjecture is essentially that these negative contributions never surpass the positive contributions provided by the regions with actual matter.

To state it in more detail and rigour, we formulate the following conjecture (with strong support from the positive energy theorem). Given the previously described initial conditions and a matter sector fulfilling the DEC, the gravitational evolution is such that the retarded integration in  
\begin{eqnarray}
h_{\mu\nu}(\bm{x},t)l^\mu l^\nu= 4\int d \bm{x}' \frac{T^{\rm T}_{\mu\nu}(\bm{x}',t')l^\mu l^\nu}{|\bm{x}-\bm{x}'|},
\label{eq:VBL-int-2}
\end{eqnarray}
with $T^{\rm T}_{\mu\nu}$ the total SET, will always be positive. We could say that the total SET satisfies a Semi-infinite Averaged B-NEC or  SAB-NEC. In other words, although the total SET contains regions with negative energies (violating the B-NEC), these violations will always be counterbalanced by positive energy contributions coming from the regions that contain the actual matter. 

If there were isolated accumulations of negative energy (as in Figure \ref{Fig:GravitationalCloudsB}) 
there could be spacetime points close to these accumulations where the negative contribution of $T^{\rm T}_{\mu\nu}$ to the gravitational potentials $h_{\mu\nu}$ could not be counteracted by positive stress-energy contributions  further away. On the contrary, with configurations in which the negative stress-energy is only screening (but not overtaking) the positive stress-energy contributions (the case of Figure~\ref{Fig:GravitationalCloudsA}), it is our conjecture that there will not exist spacetime points with negative gravitational potentials $h_{\mu\nu}$. 

In other words, our paradigm allows for the definition of a local notion of gravitational energy, distinct from (but probably somehow connected to) the Bel-Robinson definition. 
Then, we essentially conjecture that under the DEC for the matter content, the SAB-NEC or Semi-infinite Averaged B-NEC of the total SET will always be positive.

In a situation of gravitational collapse in standard general relativity, either 
\begin{enumerate}[label=\roman*),noitemsep, topsep=0pt, parsep=0pt, partopsep=0pt]
\item the $g$-cones shrink until finding a new form of stellar equilibrium (e.g. a neutron star);
\item the $g$-cones shrink until some bouncing of the structure is produced;
\item the $g$-cones shrink more and more inside the $\eta$ cones until forming a horizon and eventually some singularities. 
\end{enumerate}
In general relativity and under the DEC condition, the bouncing case ii) can occur only before any trapped region is formed (note that black-hole-to-white-hole bounce schemes of the kind described, e.g. in \cite{barcelo2011quantum,barcelo2015lifetime}, are not included in this scenario because they require a temporary violation of the DEC condition). Thus, these potential bounces refer, for example, to a star that explodes after some collapsing period, before ever forming a horizon. Here we also conjecture that in such bouncing processes, the $g$-cones expand but never exit the $\eta$-cones. In fact, sufficiently long after the explosion, when matter becomes dilute again, it is reasonable to expect that the theory reverts into a linear regime where the linear result applies.

\subsection{de Donder Harmonic Condition vs other harmonic gauge conditions}
\label{Subsec:SH-dDH}

In our present proposal, we use the dDHC to directly eliminate the linear terms on the right hand side of (\ref{NLh_equations_1}). Alternatively, one might choose to remove the linear term while simultaneously introducing additional contributions to the non-linear terms. This is for instance what the standard or Fock harmonic gauge condition does. 

We posit that a broad class of such ``harmonic conditions" could exist, all of them leading to the notion of gravity screening matter and the associated cone hierarchy that we are conjecturing. Our analysis suggests that, among this class, the dDHC stands out as a particularly elegant and minimal candidate. Notably, this approach underscores the significance of the background structure: unlike the standard harmonic condition, the dDHC inherently requires the presence of $\eta_{\mu\nu}$.

In the following section, we will explore and describe the details of our proposal based on the dDHC in the simplified context of spherical symmetry. To further contextualize our proposal, we will conclude with a comparative discussion, comparing our harmonic condition with the standard harmonic condition.

\section{Spherically symmetric stars and black holes}
\label{Sec:stars}

An example that can help to understand the proposal of this paper and its consequences is how a stellar geometry is described in terms of a $h_{\mu\nu}$ obeying the dDHC. Consider the metric exterior to a spherical distribution of static matter. After working out completely the vacuum external problem, we will say a few words about the internal stellar geometry. We begin by describing the problem at the linear level, and subsequently incorporate non-linear effects.

Recall that the linear Einstein equations in the harmonic gauge $\nabla^\mu \bar{h}_{\mu\nu}=0$ read:
\begin{eqnarray}
\square h_{\mu\nu}= -16 \pi G \bar{T}^{\rm FM}_{\mu\nu}. 
\end{eqnarray}
For a perfect fluid the SET is 
\begin{eqnarray}
T_{\mu\nu} = (\hat{\rho}+p)u_\mu u_\nu + p \eta_{\mu\nu},
\end{eqnarray}
\begin{eqnarray}
\bar{T}_{\mu\nu} = (\hat{\rho}+p) u_\mu u_\nu + {(\hat{\rho} - p) \over 2} \eta_{\mu\nu},
\end{eqnarray}
where $\hat{\rho}$ is the density of the fluid and $p$ its isotropic pressure (we write $\hat{\rho}$ to distinguish it from the coordinate $\rho$ to be introduced below).
When $p\ll\hat{\rho}$, we have 
\begin{eqnarray}
\bar{T}_{\mu\nu} \simeq  \hat{\rho} u_\mu u_\nu + {\hat{\rho} \over 2} \eta_{\mu\nu}.
\end{eqnarray}
In addition, let us consider that the fluid is at rest in the background reference frame. Then, we have
\begin{eqnarray}
\bar{T}_{\mu\nu} =  {\hat{\rho} \over 2} \delta_{\mu\nu}.
\end{eqnarray}
Finally, for simplicity, let us consider a point mass density
$\hat{\rho}=m\delta^3(\bm{x})$. Given that the problem is static, we have 
\begin{eqnarray}
\nabla^2 h_{\mu\nu} = - 8 \pi G m \delta^3(\bm{x}) \delta_{\mu\nu}.
\end{eqnarray}
By applying Gauss' law, we find the solution
\begin{eqnarray}
h_{\mu\nu} = {2 G m \over r} \delta_{\mu\nu},
\end{eqnarray}
where $r$ represents the radial coordinate of the background metric.  

In terms of a metric $\eta_{\mu\nu} + h_{\mu\nu}$ we have
\begin{eqnarray}
ds^2= -\left(1-{2 G m \over r}\right) dt^2 + \left(1+{2 G m \over r}\right) d\bm{x}^2.
\label{metric-solution-linear}
\end{eqnarray}
Notice also that the harmonic condition is satisfied. Indeed,
\begin{eqnarray}
\bar{h}_{\mu\nu} = {2 G m \over r} \delta^0_{\mu} \delta^0_{\nu};
\end{eqnarray}
and so
\begin{eqnarray}
\nabla^\mu \bar{h}_{\mu\nu} \to \partial^t \bar{h}_{tt}=0.
\end{eqnarray}
Thus, at the linear level, the radial background coordinate $r$ is precisely the radial harmonic coordinate (we call harmonic coordinates those in which the $h_{\mu\nu}$ satisfies the corresponding harmonic condition).
From the previous metric (\ref{metric-solution-linear}) and $h_{\mu\nu}$, it is clear that the null directions of the background metric $\eta$ are always outside those of the effective metric $g$.

The solution (\ref{metric-solution-linear}) is reminiscent of the Schwarzschild solution in isotropic coordinates:
\begin{eqnarray}
ds^2= -{\left(1-{G m \over 2r'}\right)^2 \over \left(1+{G m \over 2r'}\right)^2} dt^2 + \left(1+{G m \over 2r'}\right)^4 d\bm{x}^2.
\label{Schwarzschild-isotropic}
\end{eqnarray}
In fact, expanding the metric for small $Gm/2r'$, we find that  \eqref{metric-solution-linear} and \eqref{Schwarzschild-isotropic} coincide at first order if we identify $r'$ with $r$. Thus, at this linear level, the harmonic coordinate $r$ outside a spherically symmetric mass distribution is indeed very similar to the Schwarzschild isotropic coordinate $r'$. 
Let us assume for a moment that the isotropic coordinate $r'$ was precisely the radial coordinate of the background metric, $r$. Then, from the well-known relation between the radial isotropic coordinate and the Schwarzschild areal coordinate $\rho$, we would obtain the following relation:
\begin{eqnarray}
\rho = r \left(1+{Gm \over 2r}\right)^2 = r + Gm + {G^2m^2 \over 4 r}.
\end{eqnarray}
It is interesting to note that the relation between these two coordinates exhibits an offset term, $Gm$, that does not vanish as $r\to\infty$. As we will see, this feature persists in the more refined analysis that follows.

Let us now incorporate the non-linear effects in the Einstein equations. We know that the unique geometrical solution for the spherically symmetric case is the Schwarzschild solution, which in Schwarzschild coordinates reads 
\begin{eqnarray}
ds^2 = -\left(1- {2m \over \rho}\right) dt^2 + {1 \over \left(1- {2m \over \rho}\right)} d\rho^2 + \rho^2 d\Omega_2^2.
\end{eqnarray}
To correctly identify the gravitational object $h_{\mu\nu}$ that satisfies the dDHC, we need to establish an appropriate relation between this effective $\rho$ coordinate and the radial background coordinate $r$. To find this relation $\rho=\rho(r)$, we can rewrite the previous metric as
\begin{eqnarray}
ds^2 = -\left(1- {2m \over \rho(r)}\right) dt^2 + {1 \over \left(1- {2m \over \rho(r)}\right)} \left({d\rho \over dr}\right)^2 dr^2+ \rho^2(r) d\Omega_2^2
\end{eqnarray}
and look for a solution which satisfies the dDHC.

Recalling that the flat metric in polar coordinates is
\begin{eqnarray}
ds^2= - dt^2 + dr^2 + r^2 d\Omega_2^2,
\end{eqnarray}
we find that the components of the $h_{\mu\nu}$ tensor take the form
\begin{align}
&h_{00}= {2m \over \rho(r)}~; &h_{rr}={1 \over \left(1- {2m \over \rho(r)}\right)} \left({d\rho \over dr}\right)^2 -1~;
\label{hcomp1}
\\
&h_{\theta\theta} = \rho^2(r) - r^2~;   &h_{\phi\phi} =  (\rho^2(r) - r^2) \sin^2\theta~. 
\label{hcomp2}
\end{align}
Now we have to impose the dDHC condition $V_\nu=\nabla^\mu \bar{h}_{\mu\nu}=0$. A calculation assisted by Mathematica leads to a vector $V_\nu$ with only one component which is not automatically zero, namely $V_r$. Setting $V_r=0$ requires $\rho(r)$ to satisfy the following non-linear second-order differential equation:
\begin{eqnarray}
\rho''(r) + F(\rho',\rho,r)=0,
\label{DiffEq1}\end{eqnarray}
with
\begin{eqnarray}
F\left(\rho',\rho,r\right)=\frac{\rho'(r)^2}{\rho(r)\left[2m-\rho(r)\right]}+\frac{2\rho'(r)}{r}+\frac{\left[mr^2+2\rho(r)^3\right]\left[2m-\rho(r)\right]}{r^2\rho(r)^3}.
\label{DiffEq2}
\end{eqnarray}

It is easy to verify that, as expected, $\rho(r)=r$ is not a solution of the differential equation: the Schwarzschild coordinate does not satisfy the dDHC condition. The standard harmonic coordinate for a Schwarzschild geometry is known to be exactly $r=\rho-m$ (see for example the discussion in~\cite{BelinfanteGarrison1962}). The standard harmonic coordinate also fails to satisfy the previous differential equation. 

We are looking for a function $\rho(r)$ that reduces to the standard harmonic coordinate asymptotically. According to Eq.~\eqref{DiffEq2}, the behaviour of its solutions at infinity must be
\begin{eqnarray}
\rho(r) = c_1r + c_0 + \alpha(r),
\label{eq:asymexpan}
\end{eqnarray}
where $\alpha(r)$ is a radial function that vanishes at infinity and that must be obtained numerically.

We also seek for a harmonic coordinate that is in a one-to-one correspondence with $\rho$ and that extends till a putative event horizon \mbox{$\rho(r=r_H)=2m$}. It is straightfoward to verify that Eq.~\eqref{DiffEq2} is consistent with a local solution of the form
\begin{eqnarray}
\rho = 2m + a_2 \left(r-r_H\right)^2 + a_{3}\left(r-r_H\right)^3+\order{r-r_{H}}^4,
\label{eq:nearhorexp}
\end{eqnarray}
where the lowest-order coefficients are 
\begin{eqnarray}
    a_{3}=-\frac{2a_{2}}{r_{H}},\quad a_{4}=\frac{a_2}{48}\left(\frac{1}{m^2}+\frac{160}{r_{H}^2}-\frac{8a_{2}}{m}\right)
\end{eqnarray}
and $r_H$ is the horizon radius (in background coordinates).

When integrating Eq.~\eqref{DiffEq2} from the horizon outwards, we numerically search for the values of $\{r_H,a_2\}$ for which $\rho$ acquires the precise form
\begin{eqnarray}
\rho(r) \simeq r + m.
\end{eqnarray}
That is, we demand that asymptotically the coordinate $\rho$ behaves exactly like the standard Harmonic coordinate (notice that asymptotically the dDHC and the standard Harmonic condition coincide). For convenience, we can define a new function \mbox{$s(r)=\rho(r)-c_{1}r$} and solve numerically for $s(r)$ such that $\lim_{r\to\infty} s(r)=m$ (in practice, at a very large radius). Our numerical investigation is approximately consistent with the existence of this unique solution. 
We find the values
\begin{eqnarray}
    r_{H}\simeq 2.00031m,\quad a_{2}\simeq2.82717/m,
\end{eqnarray}
which yield
\begin{eqnarray}
    c_{1}-1\simeq10^{-10},\quad c_{0}-m\simeq 10^{-7}m,
\end{eqnarray}
(throughout the numerical analysis we set $m=1$).

\begin{figure}
\begin{center}
\includegraphics[scale=1]{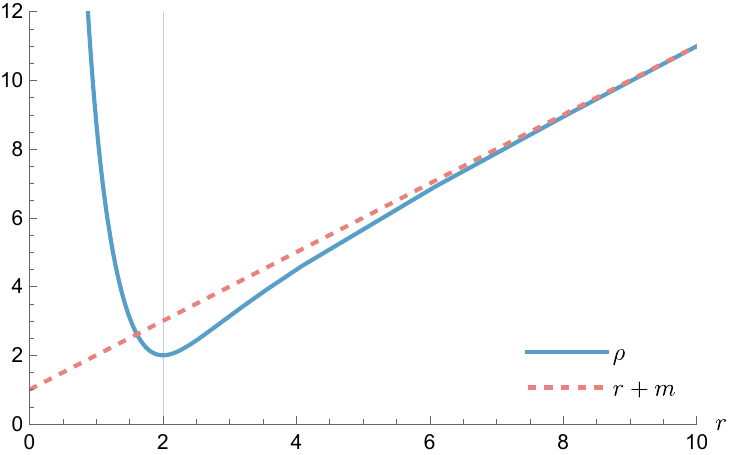}
\caption{Plot of the areal radial coordinate $\rho$ as a function of the background or harmonic radial coordinate $r$. We can appreciate how as $r\to r_{H}\simeq 2m$ (vertical line), the function $\rho$ approaches the location of a putative horizon, $\rho=2m$. 
When $r$ becomes larger, we can see that $\rho$ approaches $r+m$. Numerically we have set $G=1,~m=1$.}
\label{Fig:rho}
\end{center}
\end{figure}

To plot the full solution, we match the near-horizon solution~\eqref{eq:nearhorexp} with numerical integrations towards larger and lower values of $r$. In this way, we find a unique function $\rho(r)$ shown in Figure~\ref{Fig:rho}. The offset between $\rho$ and $r$ is approximately $m$ at infinity, and decreases as the internal static star becomes increasingly compact. In the black hole limit it leads to a behaviour of the form 
\begin{eqnarray}
\rho \simeq 2m + a_2 (r-r_H)^2, ~~~~~{\rm with}~~~~~r_H \simeq 2m,\quad a_{2}>0.
\end{eqnarray}

Now that we have obtained the function $\rho(r)$, we can calculate the form of the gravitational SET in the empty 
regions surrounding a stellar source. In this external region, we know that 
\begin{eqnarray}
-{1 \over 2}\square \bar{h}_{\mu\nu}= T^{\rm GH}_{\mu\nu}.
\label{VacuumEq}
\end{eqnarray}
Since we know the form of $h_{\mu\nu}$ (and therefore also of $\bar{h}_{\mu\nu}$) as a function of $\rho(r)$ through (\ref{hcomp1},~\ref{hcomp2}), expression (\ref{VacuumEq}) thus allows to obtain all the $T^{\rm GH}_{\mu\nu}(r)$ components simply by calculating the Laplacian of $\bar{h}_{\mu\nu}$.

\begin{figure}
\begin{center}
\includegraphics[scale=1]{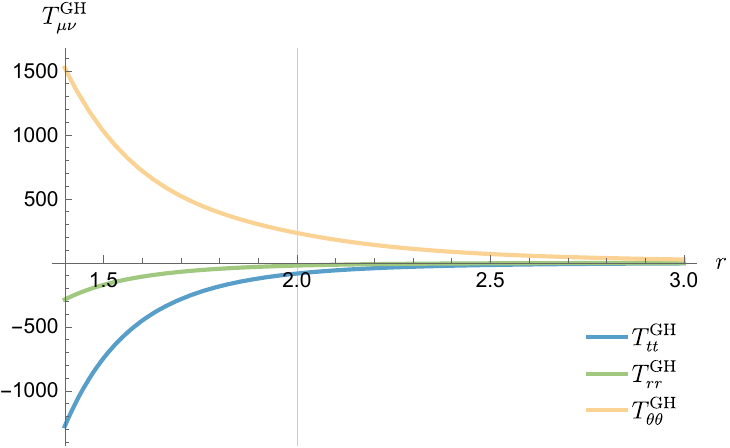}
\caption{Plot of the $T^{\rm GH}_{tt}$ $T^{\rm GH}_{rr}$ and $T^{\rm GH}_{\theta\theta}$ components of the gravitational SET as a function of the background radial coordinate $r$. We observe the negative character of this energy density. We also see the different sign of the radial and transversal pressures. All components are finite at the horizon $r=r_{H}$ (vertical line). Numerically we have set $G=1,m=1$.}
\label{Fig:SET}
\end{center}
\end{figure}

Figure~\ref{Fig:SET} shows the numerical evaluation of the different components of $T^{\rm GH}_{\mu\nu}(r)$. The explicit form of its components can be consulted in Appendix~\ref{Appendix:A}. Let us first focus on $T^{\rm GH}_{tt}(r)$. This is the gravitational energy density as such, and as expected, it is negative and increasingly negative towards the center of the configuration. The gravitational potentials generated by the central body also contribute with radial and tangential gravitational pressures, $T^{\rm GH}_{rr}$ and $T^{\rm GH}_{\theta\theta}=T^{\rm GH}_{\phi\phi}/\sin^2\theta$, respectively. As expected, these two pressures do not coincide and, in fact, have opposite signs.

The situation is analogous to what happens with the electromagnetic sector in the Reissner-Nordstr\"om solution. The Maxwell SET has components
\begin{align}
&T_{tt}^{\rm (EM)} = \frac{f(\rho) Q^2}{8\pi \rho^4}, & &T_{\hat{t}\hat{t}}^{\rm (EM)} = \frac{Q^2}{8\pi \rho^4},\\
&T_{\rho\rho}^{\rm (EM)} = -\frac{Q^2}{8\pi f(\rho) \rho^4}, & &T_{\hat{\rho}\hat{\rho}}^{\rm (EM)} = -\frac{Q^2}{8\pi \rho^4},\\
&T_{\theta\theta}^{\rm (EM)} = {T_{\phi\phi}^{\rm (EM)} \over \sin^2\theta} = \frac{Q^2}{8\pi \rho^2},
& &T_{\hat{\theta}\hat{\theta}}^{\rm (EM)} = {T_{\hat{\phi}\hat{\phi}}^{\rm (EM)} \over \sin^2\theta} = \frac{Q^2}{8\pi \rho^2},
\end{align}
where the metric function $f(\rho)$ is given by:
\begin{eqnarray}
f(\rho) = 1 - \frac{2M}{\rho} + \frac{Q^2}{\rho^2}.
\end{eqnarray}
Here, components with hats represent the contraction with a straightforward tetrad field. Associating directly the $\hat{\rho}\hat{\rho}$ and $\hat{\theta}\hat{\theta}$ component with a radial and tangential pressure, it is clear that they do not coincide, and moreover have opposite signs. 

In the gravitational situation that we are analyzing, note that we do not need to contract the components with any tetrad as the $\{t,r,\theta,\phi\}$ coordinates are already the physical coordinates, i.e. the harmonic or background coordinates. Note also that all $T_{\mu\nu}$ components are finite at the horizon limit $r\to r_H\simeq 2m$, (see again Figure~\ref{Fig:SET}). In particular, they reduce to
\begin{align}
    T_{tt}^{\rm GH}
    &=-\frac{16 m^2}{r_{H}^4}-\frac{128ma_{2}}{r_{H}^2}+\order{r-r_{H}}
    ,\nonumber\\
    T_{rr}^{\rm GH}
    &=-\frac{16 m^2}{r_{H}^4}-\frac{32ma_{2}}{r_{H}^2}+\order{r-r_{H}},\nonumber\\
    T_{\theta\theta}^{\rm GH}
    &=\frac{T_{\theta\theta}^{\rm GH}}{\sin^2{\theta}}=\frac{16 m^2}{r_{H}^2}+\frac{2r_{H}^2 a_{2}}{m}+80 m a_{2}+\order{r-r_{H}}.
\end{align}

For regular stellar-like objects with a surface located at $r_S>r_H$, the external geometry analyzed so far must be matched with the internal stellar geometry. Regardless of the internal configuration, our analysis clearly shows that the positive asymptotic mass $m$ of such a configuration consists of a matter component localized in the stellar body itself plus a negative gravitational energy surrounding the body.
The specific exterior Schwarzschild solution acquires increasingly relevant non-linear contributions as the potentials become deeper (i.e., as the star becomes more compact). These contributions include specific radial and tangential pressures in addition to a gravitational cloud of negative energy. All in all, the resulting physical $h_{\mu\nu}$ in the exterior region (i.e., using the radial harmonic coordinate) makes all null vectors of the background metric spacelike in the $g$-metric. This can easily be seen for radial null vectors since both $h_{00}$ and $h_{rr}$ are greater than zero (the same happens for Schwarzschild coordinates). A representative tangential null vector of the background metric has the form $l^{\mu}=\{1,0,1/r,0\}$. Then, one can calculate $l^\mu l^\nu h_{\mu\nu}$ yielding
\begin{equation}
h_{00} +{1 \over r^2} h_{\theta\theta}= {2m \over \rho(r)} + {\rho^2(r) -r^2 \over r^2}.
\end{equation}
For example, at the horizon limit, this quantity results in $2m/\rho(r_H) \sim 1 >0$.
In the asymptotic region, we instead have
\begin{equation}
{2m \over r-m} + {(r-m)^2 -r^2 \over r^2} \simeq {m^2 \over r^2} >0.
\end{equation}

Now, Eq. (\ref{eq:VBL-int-2}) tells us in turn that any integration along the causal past light cone of the background metric results in a positive SAB-NEC. 
For reasonable stellar interiors (i.e. with positive-energy densities), the effective cones will continue to stretch inside the background cones as we move towards their interiors. 
For regular stars, this process would always end with some minimal-angle cone (i.e. not a degenerate cone) at the radial origin. 
Therefore, the causal cones of the effective metric in the internal region would also be inside those of the background metric.
Thus, for reasonable interiors, we find full consistency with our general conjecture.

The situation is equivalent to what happens with the Newtonian potential of a spherically symmetric stellar-like object. The potential grows from some finite negative value at the centre towards another negative value at the surface. Then it continues decreasing as $-m/r$ until reaching zero at infinity. It is interesting to mention that in the case of a matter shell with an empty interior, we will find that the effective metric in the interior is Minkowski. However, this Minkowski metric is not the same as the background Minkowski metric. The HBP allowed us to give meaning to the previous sentence. We have ``the speed of light", as controlled by the real background metric, and a ``slowed-down speed of light" inside the matter shell.

In principle, one could build up configurations as close to black hole compactness as desired. However, to surpass the Buchdahl and Bondi compactness limits \cite{Buchdahl:1959zz,Bondi:1964zz} one would need to introduce either huge matter anisotropic pressures, violating the DEC, or directly introduce additional negative energy sources, which will clearly violate all energy conditions (see the discussion in~\cite{Arrechea:2024vxp}). 
We leave for future work to analyze in detail whether such non-standard configurations can also be accommodated within our HBP. But it is already interesting to speculate on what would happen if the black hole limit is reached.

From a geometric point of view the $r$ coordinate is non-regular at the horizon. This is so because the determinant of the effective metric would become zero at a horizon: 
\begin{equation}
    \sqrt{-g}|_{r\to r_H} = \left({d\rho \over dr} \right) \rho^2(r) \bigg|_{r\to r_H} \sin\theta \simeq  8m^2 a_2(r-r_H) \sin\theta \to 0.
\end{equation}
The same happens in isotropic coordinates. From the point of view of our HBP, 
the existence of an eternal black hole configuration would signal a breakdown of this causal-deformation theory of gravitation.
In static configurations, the (unique) effective cones are symmetrically stretched inside the Minkowski cones. The black hole limit corresponds to a situation in which the cone degenerates, leading to a zero effective speed of light: any perturbation will be frozen when a horizon appears, and gravity could no longer be considered a causality deformation. We expect that in this circumstance, an appropriate description of the physics would require incorporating new terms beyond those considered by the GR low-energy description. 

\begin{figure}
\begin{center}
\includegraphics[scale=0.7]{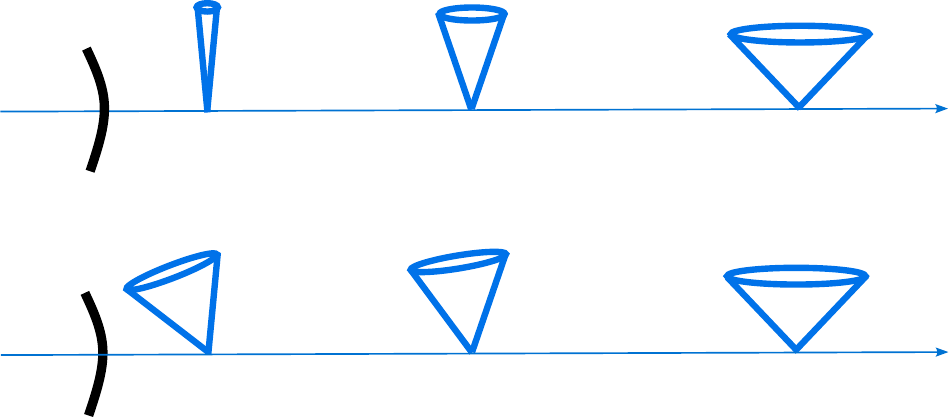}
\caption{Pictorial representation of how the effective causal cones in the radial direction behave as one approaches a horizon surface (in black on the left). The upper figure represents the behaviour of the cones with the radial distance for the static configurations analyzed in this paper. The lower figure is a representation of what we expect to happen with the cones in the dynamical formation of a horizon. }
\label{Fig:stat-vs-dyn-cones}
\end{center}
\end{figure}

Notice, however, that we do not expect this same breakdown of the effective description to occur when dealing with the dynamical generation of black holes in a collapse situation. The unique potentials describing such a situation will not be the same as the ones we have found for static configurations. We expect that, when reaching the black hole limit in a dynamical configuration, the effective cones will not only stretch but also tilt. In this way, one can produce a horizon without breaking the effective theory, i.e. without producing a singular cone. We illustrate the difference in behaviour between static and dynamic cones in the black hole limit in Figure~\ref{Fig:stat-vs-dyn-cones}. An argument in favour of this expectation can already be seen at the linear level: as the SET source in collapsing scenarios would have a non-diagonal $tr$ term, it would lead to distinct gravitational potentials $h_{\mu\nu}$. 
Recall that in the Geometrical paradigm these different $h$'s are just related by gauge transformations, so the difference is not appreciated. In other words, a relational account of low-energy gravitational phenomena, based exclusively on the effective metric, would not be able to distinguish between a static and a collapsing configuration. The HBP offers new perspectives which might be useful when going beyond low-energy regimes. In this regard, it is also interesting to realize that semiclassical physics points to similar ideas: semiclassical gravity is not consistent with the existence of eternal and static horizons, but it is perfectly consistent with horizons formed by collapse which start evaporating as soon as they have formed~\cite{Arrechea2024}. 

Let us finish this section by comparing our dDHC radial coordinate $\rho(r)$ with the standard Harmonic coordinate $\rho(r)=r+m$. This standard Harmonic coordinate can be written as $\rho=2m + a_{1}(r-r_H)$, with $r_H=m$. We can calculate the behaviour that the corresponding $T^{\rm GH}_{tt}$ would have close to the horizon in terms of this coordinate. In the black hole limit, we obtain 
\begin{align}
    T^{\rm GH}_{tt} 
    &
    =- \frac{4m a_{1}}{\left(r-r_{H}\right)^3}+\order{r-r_{H}}^{-2},\nonumber\\
    T^{\rm GH}_{rr} 
    &
    =- \frac{4m a_{1}}{\left(r-r_{H}\right)^3}+\order{r-r_{H}}^{-2},\nonumber\\
    T^{\rm GH}_{\theta\theta} 
    &
    =\frac{T^{\rm GH}_{\phi\phi}}{\sin^{2}{\theta}}= \frac{4m r_{H}^2a_{1}}{\left(r-r_{H}\right)^3}+\order{r-r_{H}}^{-2}.
\end{align}
In this case, the gravitational energy density would therefore diverge in the black hole limit. From this perspective, the dDHC coordinate produces a softer behaviour at the horizon limit than the standard Harmonic coordinate. The dDHC yields a picture similar to what happens to the electric energy in the Reissner-Nordstr\"om solution, as already mentioned before. Indeed, let us consider a charged ball of mass with a size approaching the would-be horizon. The electric field and its associated energy density remain finite as the horizon is approached because of the extended nature of the charge distribution. The same thing happens with our gravitational energy density: it does not diverge at the black hole limit.

The dDHC is probably not the unique choice leading to a cone hierarchy which preserves a more fundamental background causality, but as far as we can see it is the simplest guess which exhibits appealing properties such as the ones described so far.

\section{Conclusions and further discussion}
\label{Sec:conclusions}

In a previous work~\cite{BarceloJannes2024} we introduced a novel paradigm with which to understand gravitational phenomena: we called it the Harmonic Background Paradigm (HBP). In this work we have further elaborated its presentation by analyzing its non-linear aspects.

The HBP allows understanding gravity as a deformation of a deeper underlying causality. We take this deeper causality to correspond to a simple flat Minkowski structure. Moreover, we provide conditions under which it is conjectured that the effective causality associated with gravity will always be subsumed into the Minkowskian causality: we write $[g]\leq [\eta]$. Satisfying this cone hierarchy condition is necessary if one wants to give a proper ontological role to a background causality. 

Within the HBP, the presence of this background causality allows for a simple and rigorous definition of a gravitational stress-energy tensor representing the local gravitational energy.
Then, the central contribution of this work is to identify a connection between the presence of a causal-cone hierarchy and a positivity of the matter energy content. 
The idea is that, assuming that any matter source has a positive energy contribution, it sparks the generation of pure-gravity negative-energy clouds. However, these negative-energy clouds never overshadow the positive energy of the initial matter seeds. Then, the total gravitational effect at any point in spacetime is to generate an effective causality which is always inside the background causality. In fact, the idea is a reincarnation of the positive energy theorem of GR, but now using a strictly local definition of gravitational energy.

In the present paper, we have discussed how some standard behaviours can be analyzed from the HBP perspective, leaving the analysis of specific beyond-GR situations for future work. In particular, we have analyzed how the HBP applies to spherically symmetric configuration sourced by a central and static matter content. We have found the relation between the Schwarzschild radial coordinate $\rho$ and a background or ``harmonic'' radial coordinate $r$. Interestingly, the relation between $\rho$ and $r$ is quadratic near a would-be horizon $\rho \simeq 2m+ a_2(r-r_H)^2$, and linear $\rho \simeq r+m $ for $r\to +\infty$. An observer with access to both the effective and the background metrics would associate a different area to a sphere of radius $r$: they would obtain $4\pi \rho^2$ using effective measurements and $4\pi r^2$ using background measurements. At large radii the relative difference is negligible, 
$(\rho^2-r^2)/r^2 \simeq 2m/r$, but in principle those observers could use this difference to measure how much mass the spacetime contains.

An additional interesting consequence is the following. Due to the appearance of the negative gravitational energy clouds around any central object, the total mass-energy contained within a sphere of radius $r$ around this central object will be
\begin{equation}
m_l(r)=4\pi \int_0^r r'^2 dr' [\rho_{\rm M}(r)+\rho_{\rm G}(r)]~.
\end{equation}
This total mass-energy
diminishes with the radius $r$, as more and more negative energy is encompassed. Asymptotically it reaches the ADM value $m$. This mass notion is, therefore, distinct from the Misner-Sharp mass, which for a Schwarzschild exterior geometry is just the constant \mbox{$m_{\rm MS}(r)=m$}. Our conceptualization makes the understanding of the Schwarszchild solution closer in spirit to the GR treatment of the Reissner-Nordstr\"om solution: both the Coulomb energy of a charge and the Coulomb energy of a mass are distributed around the location of the source.

At this point, let us also comment on a historical curiosity. When Einstein received Schwarzschild's manuscript with the solution of the Einstein equations in spherical symmetry, he expressed surprise about the simplicity of the solution: ``I did not expect that one could formulate so easily the rigorous solution to the problem"~\cite{CPAE8}. 
Einstein's expectation was obviously due to the complicated non-linear character of his equations and makes perfect sense, as our treatment of the spherically symmetric solution based on the dDHC condition illustrates. 
Indeed, the gravitational potential energy associated with a central mass itself gravitates. Taking into account these self-interactions, the overall potential becomes more complicated than a straightforward ``Coulomb gravitational potential" $-m/r$. It is thus indeed the complex non-linearity of this problem that has only allowed us to find a numerical solution.

If the historical route had been different, starting from the HBP perspective, then realizing that there is a twinned gauge formulation of the gravitational phenomena allows one to bring about all the flexibility and formal power of using different gauges. Remarkably enough, the Schwarzschild gauge with its area radial coordinate allows one to find a simple form for the spherically symmetric vacuum metric. This illustrates how different paradigms can also help each other in developing a corpus of knowledge.

If the conjecture presented in this work is correct, we have at our disposal a third paradigm to understand gravitational phenomena. It would be most interesting to re-examine other standard GR behaviours in light of this novel paradigm. It would also provide additional perspectives to the historical and philosophical debate regarding the ``hole argument'' (see e.g.~\cite{Stachel2014}).
However, the full power of a new paradigm would only crystallize when analyzing situations that push general relativity to its limits, such as the onset of the formation of singularities. 
In this sense, although we have not analyzed such situations in the present work, the HBP is ready to incorporate deviations from the GR behaviour based on the availability of a second causality. For example, given that our proposal treats gravity as a theory without gauge symmetries, it smooths out many of the crucial obstacles to recover general relativity as an emergent feature from underlying systems with a standard notion of energy and locality~\cite{Marolf:2014yga}. More generally, describing gravity as a bi-metric theory allows, among many other things~\cite{BarceloJannes2024}, to separate the idea that $h_{\mu\nu}$ might exhibit some quantum behaviour from that of quantizing the entire spacetime structure\footnote{We are not claiming that a full dissolution of space and time is not going to be necessary at a deeper level of description. We just believe that it might be useful to proceed step by step.}. The presence of an explicit background could help to solve many of the problems associated with finding a quantum gravity theory, without giving up or contradicting the great insights that GR has brought us.

\begin{acknowledgements}
The authors would like to thank Valentin Boyanov, Eric Curiel, Gerardo Garc\'{\i}a-Moreno, Marc Mars, Carlos F. Sopuerta and Matt Visser for enlightening discussions on the subject.
Financial support was provided by the Spanish Government through the Grant
PID2023-149018NB-C43 
(funded by MCIN/AEI/10.13039/501100011033), and by the Junta de Andaluc\'ia through the project FQM219. CB acknowledge financial support from the Severo Ochoa grant CEX2021-001131-S funded by MCIN/AEI/10.13039/501100011033.
\end{acknowledgements}

\appendix
\section{Gravitational stress-energy tensor}
\label{Appendix:A}

In this Appendix, we show the explicit expressions of the gravitational stress-energy tensor components:
\begin{align}
    T_{tt}^{\rm GH}
    &
    =\frac{2 \rho  \left(\rho ''\right)^2}{2 m-\rho }-\frac{4 m \left(\rho '\right)^4}{(\rho -2 m)^3}+\rho ' \left(\frac{2 \rho  \rho ^{(3)}}{2 m-\rho }+\frac{4 \rho  (2 m-\rho ) \rho ''}{r (\rho -2 m)^2}+\frac{4 m}{\rho ^2 r}+\frac{8 \rho }{r^3}\right)+\rho '' \left(\frac{2 m}{\rho ^2}-\frac{4 \rho }{r^2}\right)\nonumber\\
    &
    +\left(\rho '\right)^2 \left(-\frac{4 m}{\rho ^3}+\frac{10 m \rho ''}{(\rho -2 m)^2}-\frac{4}{r^2}\right)+\frac{4 m \left(\rho '\right)^3}{r (\rho -2 m)^2}-\frac{4 \rho ^2}{r^4},\nonumber\\
    T_{rr}^{\rm GH}
    &
    =\left(\rho '\right)^2 \left(\frac{10 m \rho ''}{(\rho -2 m)^2}-\frac{4 \left(-2 m^2 r^2+2 m \rho ^3+m \rho  r^2-3 \rho ^4\right)}{\rho ^3 r^2 (\rho -2 m)}\right)-\frac{2 \rho  \left(\rho ''\right)^2}{\rho -2 m}-\frac{4 m \left(\rho '\right)^4}{(\rho -2 m)^3}\nonumber\\
    &
    +\frac{4 \rho ^2 (2 m-\rho )^3}{r^4 (\rho -2 m)^3}-\frac{2 (2 m-\rho )^3 \rho '' \left(m r^2+2 \rho ^3\right)}{\rho ^2 r^2 (\rho -2 m)^3}\nonumber\\
    &
    -\rho ' \left(\frac{2 \rho  \rho ^{(3)}}{\rho -2 m}+\frac{4 (2 m-\rho )^3 \left(m r^2-2 \rho ^3\right)}{\rho ^2 r^3 (\rho -2 m)^3}+\frac{4 \rho  \rho ''}{r (\rho -2 m)}\right)+\frac{4 m \left(\rho '\right)^3}{r (\rho -2 m)^2},\nonumber\\
    T_{\theta\theta}^{\rm GH}&
    =\frac{T_{\phi\phi}^{G}}{\sin^2{\theta}}=\frac{2 \rho  r^2 \left(\rho ''\right)^2}{\rho -2 m}+\frac{2 m r^2 \rho ''}{\rho ^2}+\frac{4 m r^2 \left(\rho '\right)^4}{(\rho -2 m)^3}\nonumber\\
    &
    +\left(\rho '\right)^2 \left(\frac{4 \rho  (2 m-\rho )}{(\rho -2 m)^2}+\frac{4 m r^2 (2 m-\rho )}{\rho ^3 (\rho -2 m)}+\frac{10 m r^2 (2 m-\rho ) \rho ''}{(\rho -2 m)^3}\right)\nonumber\\
    &
    +\rho ' \left(\frac{2 \rho  \rho ^{(3)} r^2}{\rho -2 m}+\frac{4 \rho  r \rho ''}{\rho -2 m}+\frac{4 m r}{\rho ^2}\right)-\frac{4 m r \left(\rho '\right)^3}{(\rho -2 m)^2}+\frac{4 \rho ^2}{r^2}.
\end{align}


%

\end{document}